\documentclass[10pt]{article}
\usepackage{amssymb, amsmath, verbatim, epsfig,setspace,mathtools}
\usepackage{hyperref}

\newcommand{\fig}[1]{{\bf Fig.\ \ref{#1}}}

\title{\bf 
\vspace{-1in}
\LARGE{Functionally graded keratin facilitates\\ tactile sensing in elephant whiskers}
}
\date{\vspace{-3ex}} 
\author{Andrew K. Schulz$^{1,*}$, Lena V. Kaufmann$^{2,+}$, Lawrence T. Smith$^{3,+}$,\\Deepti S. Philip$^{1,4}$, Hilda David$^{4}$, Jelena Lazovic$^{5}$,\\Michael Brecht$^{2}$, Gunther Richter$^{4}$, Katherine J. Kuchenbecker$^{1,*}$\\
\small{$^{1}$Haptic Intelligence Department, Max Planck Institute for Intelligent Systems (MPI-IS), Stuttgart, Germany}\\
\small{$^{2}$Bernstein Center for Computational Neuroscience Berlin, Humboldt University of Berlin, Berlin, Germany}\\
\small{$^{3}$Robotic Materials Department, MPI-IS, Stuttgart, Germany}\\
\small{$^4$Materials Central Scientific Facility, MPI-IS, Stuttgart, Germany}\\
\small{$^5$Medical Systems Central Scientific Facility, MPI-IS, Stuttgart, Germany}
}

\begin{document}
\maketitle
\noindent\text{\small{$+$ denotes equal contribution}}\\ 
{\bf * Co-corresponding authors:}\\
Andrew K. Schulz and Katherine J. Kuchenbecker\\
Heisenbergstraße 3, 70569 Stuttgart, Germany\\
aschulz@is.mpg.de \& kjk@is.mpg.de

\doublespacing 
\subsection*{Abstract} 

Keratin composites enable animals to hike with hooves, fly with feathers, and sense with skin. These distinct functions arise from variations in the underlying properties and microscale arrangement of this natural polymer. One well-studied example is mammalian whiskers, elongated keratin rods attached to tactile skin structures that extend the animal's sensory volume. Here, we investigate the non-actuated whiskers that cover Asian elephant (\textit{Elephas maximus}) trunks and find they are geometrically and mechanically tailored to facilitate tactile perception by encoding contact location in vibrotactile signal amplitude and frequency. Elephant whiskers emerge from armored trunk skin and shift from a thick, circular, porous, stiff root to a thin, ovular, dense, soft point. This smooth transition enables interaction with widely varying substrates, reduces wear, and increases the vibrotactile signal information generated during contact. The functionally graded geometry, porosity, and stiffness of elephant whiskers tune the neuromechanics of trunk touch, facilitating highly dexterous manipulation.  

\doublespacing 

\subsection*{Introduction} Whiskers, tree bark, seashells, and other biological composites consist of proteins, sugars, and minerals~\cite{eder_biological_2018}, giving living organisms diverse mechanical and functional properties through morphology and composition~\cite{wang_keratin_2016}. One of these biological building blocks, the protein keratin, is arranged in alpha helices in mammalian epidermis, horns, nails, and hair, such as the dense, hollow fur that helps polar bears thermoregulate in arctic temperatures~\cite{wu_biomimetic_2023}. Alpha-keratin helices occur at the 7-nm length scale~\cite{squire_fibrous_1987}, and the structures they create have diverse mechanical forms that support the keratinized structure's functional use: for example, bighorn sheep horns contain hollow tubules that increase shock absorption~\cite{huang_hierarchical_2017}, and the tapering of rat whiskers facilitates texture discrimination by allowing the whisker points to mechanically interact with tiny surface asperities~\cite{williams_advantages_2010}. The thicker root of each rat whisker is actuated by a collagen protein wrapper, increasing the animal's sensory volume~\cite{grant_what_2022}. Keratin itself cannot sense touch, but this keratin-collagen interface is ringed by sensory neurons, creating the three components of familiar whiskers: hair, collagen wrapper (follicle), and sensory ring (\fig{fig1}A)~\cite{staiger_neuronal_2021}. 

This intriguing somatosensory structure has both incentivized biologists and neuroscientists to understand whisker neuromechanics~\cite{voges_structural_2012,ginter_comparative_2011,hans_mechanical_2014,laturnus_functional_2021,oladazimi_biomechanical_2018} and inspired several robotic whisker designs~\cite{zhao_novel_2022,gul_fully_2018,valdivia_y_alvarado_design_2012, solomon_robotic_2006}. Past research on whisker morphology and function has primarily focused on the follicle and sensory ring structures~\cite{kim_mechanism_2012,deiringer_functional_2023,takatoh_vibrissa_2018}, as well as the scaly wall known as the cuticle (\fig{fig1}B). Mouse whiskers increase in elastic modulus from the root to the point by a factor of two~\cite{quist_variation_2011}, and seal whiskers decrease in modulus by the same ratio~\cite{kamat_undulating_2024}. Researchers typically average these moduli~\cite{zhao_novel_2022,oladazimi_conveyance_2021,oladazimi_biomechanical_2018} and assume the whisker is solid and uniform throughout. 
Instead of exploring the effects of nonuniform material properties, the focus has been on morphology, determining how whisker length~\cite{voges_structural_2012}, tapering~\cite{williams_advantages_2010}, aspect ratio~\cite{kamat_undulating_2024}, and actuation~\cite{grant_active_2009} impact the mechanical frequencies or torques communicated to the whisker root (\fig{fig1}B). Elephant trunk whiskers are unique among studied mammalian taxa as they lack muscles in the root follicle (\fig{fig1}C)~\cite{deiringer_functional_2023}. These non-whisking whiskers adorn a hydrostat with approximately 90\,000 muscle fascicles~\cite{longren_dense_2023} capable of near-infinite-degree-of-freedom motions~\cite{schulz_skin_2022} that generate ample whisker-object contact. 

Whiskers extend the sensory function of the elephant trunk, assisting with precision manipulation of widely varying items including hundreds of kilograms of food daily~\cite{deiringer_functional_2023,schulz_suction_2021}. We hypothesize that the elephant's lack of individual whisker actuation is accompanied by striking geometrical and material differences compared to actuatable whiskers, such as those of domesticated cats (\textit{Felis catus}) or rats (\textit{Rattus norvegicus}). This hypothesis is tested in baby and adult Asian elephants (\textit{Elephas maximus}) by studying the three parameters governing their mechanical behavior: geometry, porosity, and elastic modulus~\cite{hartmann_vibrissa_2016} (\fig{fig1}E). Microscopy and material characterization allow us to describe how elephant whisker parameters shift from root to point. Finite element analysis (FEA) of whiskers with nonuniform cross section, porosity, and modulus demonstrates how whisker morphology and composition affect what is felt by the sensory neurons at the whisker root.

\subsection*{Tapered ellipsoidal whiskers highlight contact directionality}
\begin{figure}[thbp]
      \centering
\includegraphics[width=1\textwidth,page=1,trim=0.0in 0in 0in 0.0in,clip]{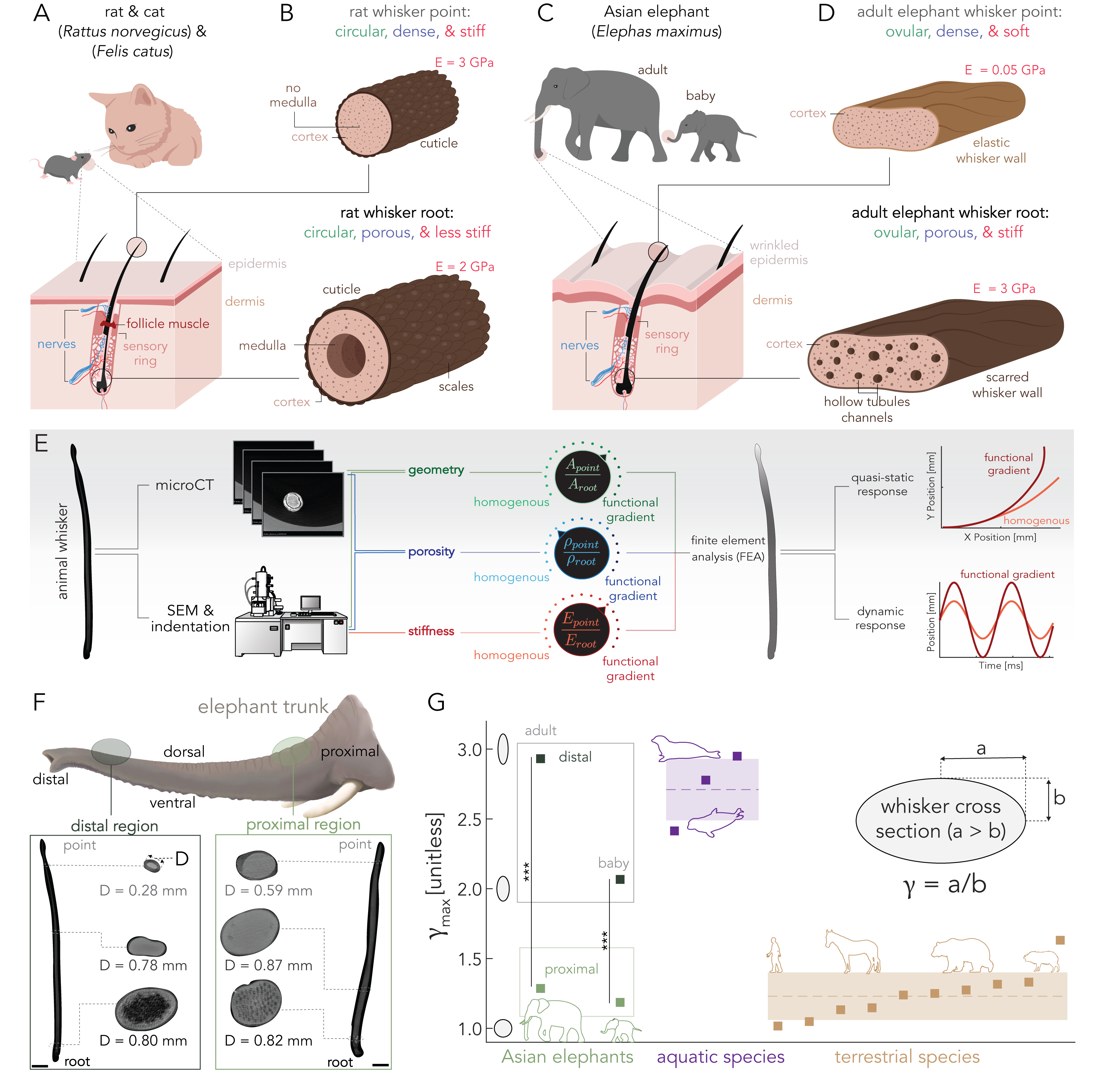}
      \caption{\label{fig1}
      \textbf{Morphology of animal whisker systems with a framework for studying whisker geometry, porosity, and stiffness.} 
      (A) The rat whisker-skin system with whisker morphology shown in (B) displays differences between root and point. 
      (C) The Asian elephant whisker-skin system showing distal trunk whiskers, which lack a follicle muscle~\cite{deiringer_functional_2023}, with whisker morphology shown in (D) displaying differences between root and point. 
      (E) Experiment flowchart used to study how the three variables of whisker geometry, porosity, and stiffness affect tactile perception through the whisker. 
      (F) Schematic of an elephant trunk showing the two regions (distal and proximal) from which whiskers were analyzed, with an enlarged microCT rendering and three cross sections for each representative whisker. 
      (G) The maximum aspect ratio, $\gamma_{\max}$, of Asian elephant trunk whiskers compared to whiskers of aquatic and terrestrial species, with elephants displaying significant differences in cross-section geometry between distal and proximal regions $(^{\star\star\star} ~p < 0.001)$.
      }
\end{figure}

We used microscopy and microCT to characterize elephant whisker geometry, including longitudinal tapering and cross-sectional shape. First, whiskers from the distal and proximal sections of an Asian elephant's trunk were compared. We imaged each whisker from its root (embedded in the skin) to its point (extending beyond the skin and interacting with the environment). Whiskers from the distal trunk are thin, highly tapered, and blade-like (\fig{fig1}C--F, fig.\ S1A--B, movie S1). The whisker cross section's long axis aligns with the trunk's circumferential wrinkles (fig.\ S1D), and its ellipsoidal aspect ratio increases by 40\% as the elephant develops (\fig{fig1}G). The whiskers of pinniped species also have an ovular cross-sectional shape (\fig{fig1}G, table S1), which confers sensory benefits in preferred directions in water~\cite{kamat_undulating_2024}. The distal trunk is used for gripping and manipulation~\cite{dagenais_elephants_2021}, so its blade-like whiskers could facilitate directional contact perception; indeed, researchers have long hypothesized that elephants use their distal whiskers for precise tactile sensory discrimination~\cite{dehnhardt_sensitivity_1997}. This blade-like whisker structure also appears more rectangular than ovular near the root, which could alter the bending rigidity in preferred directions similar to the round-to-square cross-sectional transition in bird feather rachis~\cite{wang_light_2017}.

Whiskers from the proximal trunk are thicker, only gently tapered, and nearly circular (\fig{fig1}F, fig.\ S1C, movie S1). Their aspect ratios match those of other terrestrial animals' whiskers (table S1) and are significantly more circular than the whiskers of the distal trunk and aquatic mammals ($n=5$, \fig{fig1}G, $p< 0.001$). Elephant proximal whiskers are also wavy, with undulating radius changes along the whisker length (\fig{fig1}F, fig.\ S1C, movie S1), similar to pinnipeds~\cite{kamat_undulating_2024}, which could reduce vibrations when moving through air or water, as has been shown in seals~\cite{kamat_undulating_2024}. Interestingly, proximal elephant trunk whiskers are much less tapered than distal whiskers, and their points round to a diameter of 0.59 $\pm\ 0.07$\ mm ($n=5$, \fig{fig1}F). Similar to mechanical curb feelers on early cars, the elephant's proximal whiskers may perform omnidirectional proximity sensing to identify obstacles near the face, compensating for the animal's poor eyesight~\cite{garstang_elephant_2015}. 

Considering each whisker as a long, slender rod, we modeled its flexural rigidity, $K$, as a function of the material stiffness, $E$, and the cross section's second moment of area, $I$~\cite{cai_fundamentals_2016}. Therefore, whisker tapering, or reduction of the diameter $D$ along the whisker length $z$, represents an influential method of controlling local whisker flexibility, as $K \propto D^4$. (Graded material stiffness, $E(z)$, is explored later.) A tiny diameter at the point (as seen in distal but not proximal elephant trunk whiskers) helps with texture discrimination~\cite{hires_tapered_2013} and allows penetration into the depth of rough materials to detect surface features~\cite{jadhav_sparse_2009} and even discriminate shapes~\cite{hires_tapered_2013}. These geometrical parameters of aspect ratio and tapering are the current focus of many whisker researchers, while few studies have explored micron-scale factors that affect whisker mechanics. 

\subsection*{Elephant whiskers exhibit a horn-like microstructure}
\begin{figure}[thbp]
      \centering      \includegraphics[width=1\textwidth,page=1,trim=0.0in 0in 0in 0.0in,clip]{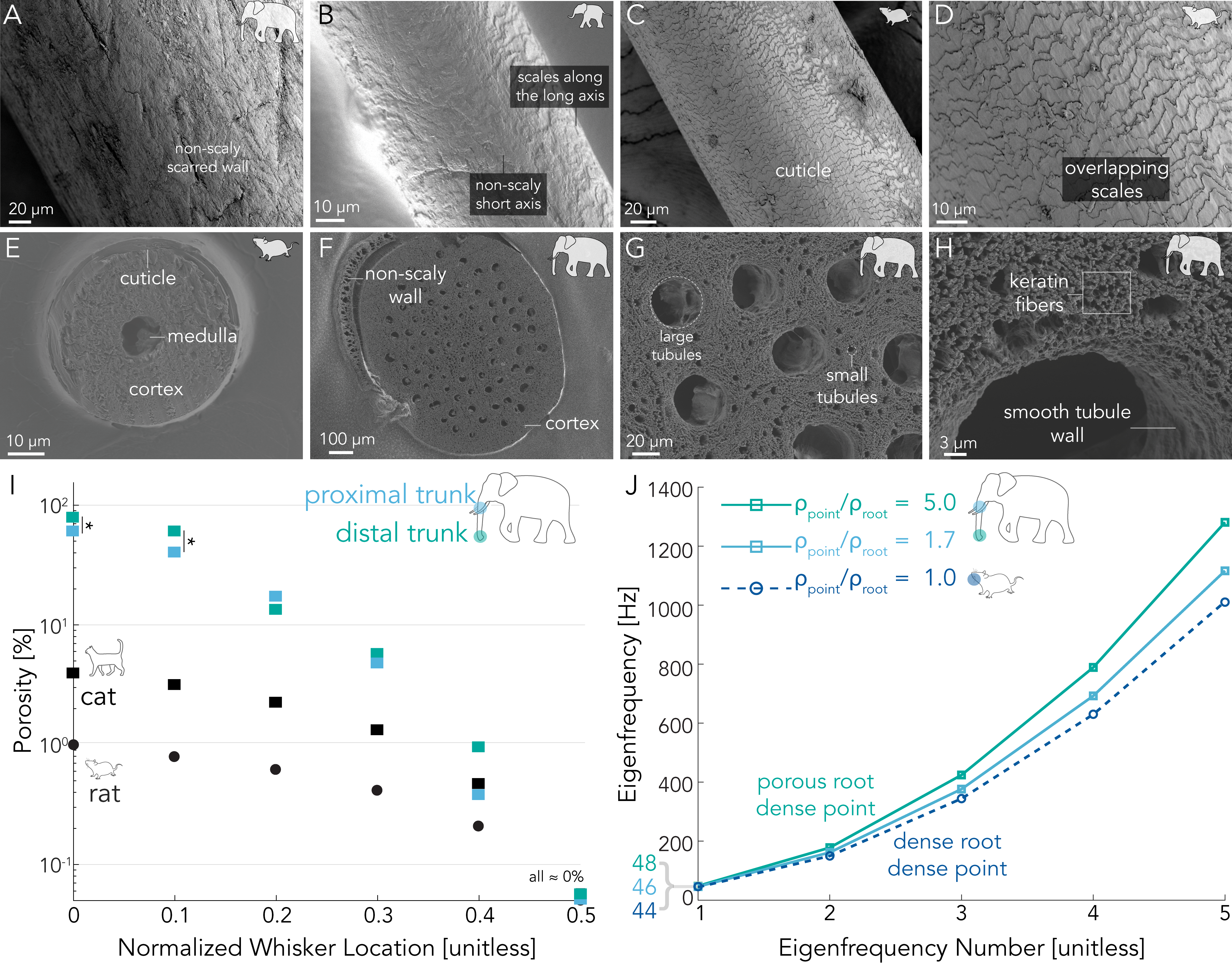}
      \caption{\label{fig2}
      \textbf{Elephant whiskers display horn-like microstructure that decreases mass and increases vibration frequency.} 
    SEM images of various whisker wall structures with adult elephant whisker root in (A), adult elephant distal whisker root in (B), baby elephant distal whisker root in (C), and rat whisker root in (D). SEM images of transverse whisker cross sections with rat whisker shown in (E), elephant whisker cortex and non-scaly wall in (F), tubule channels in elephant whisker root in (G), and individual longitudinal keratin fibers around tubule wall in (H). 
    (I) Porosity of four different whisker types (rat, cat, elephant distal trunk, elephant proximal trunk) along the normalized whisker location. Data of rat whiskers taken from Voges et al.~\cite{voges_structural_2012}. Significant differences were found between proximal and distal adult elephant whiskers at the root ($^\star ~p < 0.05$). 
    (J) FEA outputs of the free-vibration response of whiskers with three different density ratios between the point and root. Simulated whiskers with a porous root and dense point have eigenfrequencies that are spread across a broader range of frequencies than otherwise identical whiskers with constant density. 
      }
\end{figure}

Scanning electron microscopy (SEM) images showed that most Asian elephant whiskers analyzed from the adult (both distal and proximal sections) are encased by an axially scarred cuticle (\fig{fig2}A) that appears to be unique among studied whiskers. In contrast, portions of the roots of baby elephant whiskers and all examined elephant body hairs have a scaly cuticle (fig.\ S2A) like that found along the lengths of whiskers and body hairs of other terrestrial mammals (\fig{fig2}B--D, fig.\ S2A,C). The primary function of a scaly cuticle is thought to be to preserve the mechanical integrity of the hair during deformations~\cite{yang_strength_2020}. The baby elephant distal trunk whiskers that we imaged had scales only at the widest points of the root; thus, these scales might contribute to development of the whisker as a blade-like structure, as the scales grow over time (fig.\ S1B). However, there is no indication of scales at the baby elephant whisker point (fig.\ S2B), indicating they do not wear off later from abrasion, which is commonly assumed when whiskers lack a scaly cuticle~\cite{dougill_describing_2023}. Given this lack of scales and the scarred outer wall of elephant trunk whiskers, the most similar keratin structures are mammalian horns, antlers, and porcupine quills~\cite{yang_axial_2013}. Such structures contain porous cavities, known as tubules, along their length. 

We observed porous macro- and micro-scale axial channels in the microCT images of elephant trunk whisker roots (\fig{fig1}F, movie S1). These channels were visualized by SEM imaging cryotome-sliced whisker sections showing the transverse cross section (fig.\ S3A--B). The inner matrix of elephant whiskers has longitudinal keratin strands packed and wrapped tightly around large and small tubules (\fig{fig2}F--H). For comparison, the tail hair of Asian elephants has been shown to contain a few medulla channels~\cite{yates_forensic_2010}, and elephant body hair has a single medulla channel~\cite{yang_strength_2020}. The macro channels seen in elephant whiskers are 30--40 $\mu$m in diameter (\fig{fig2}F-G) and are arranged at the root, similar to the tubules in antlers, bighorn sheep horn (fig.\ S4A-B), and horse hooves~\cite{mi_creating_2019}. Tubules provide added benefits for energy dissipation in horn structures~\cite{yang_axial_2013} and change the material porosity only slightly~\cite{mi_creating_2019}. Therefore, the more numerous tubule channels present in elephant whiskers may provide benefits beyond the energy dissipation of previously studied keratinous horns. 

\subsection*{Functionally porous whiskers amplify tactile signal transmission through mass reduction}
We determine the porosity along the length of each studied elephant whisker using microCT and see a strong porosity gradient: elephant whisker roots are highly porous ($\sim$70\%) and transition from $\sim$15\% porosity at 20\% of normalized whisker length to fully dense ($\sim$0\% porosity) by the middle of the whisker. Distal trunk whiskers have a root porosity of 82 $\pm$ 9\% ($n=5$) with the numerous medullary channels appearing as a hollow shell (\fig{fig1}G, \fig{fig2}I). Proximal whiskers have a somewhat less-porous root (63 $\pm$ 7\%, $n=5$) that also promptly transitions to a dense structure (\fig{fig2}I); the density difference between the roots of proximal and distal trunk whiskers is significant ($p < 0.05$). Since the medullas of cat and rat whiskers occupy only 2--5\% of the whisker cross section (\fig{fig2}I)~\cite{voges_structural_2012}, previous investigations have assumed that all whiskers are fully dense, but elephant whiskers clearly break this assumption.

A whisker transmits tactile information from contacts along its length to the sensing structures at its root through vibrations, enabling the animal to distinguish obstacles, food, or prey~\cite{diamond_where_2008}. Experiments on whisker models with axially graded porosity (table S2, fig.\ S5) indicate that functional grading of porosity plays a role in transmitting this information. Finite element analysis of graded-porosity whiskers shows that elevated root porosity modestly increases the natural frequency of the first five vibration modes ($<$10\%) compared to whiskers with no porosity (\fig{fig2}J). The increase in frequencies follows intuitively from basic beam theory~\cite{cai_fundamentals_2016}, as porous whiskers have reduced volumetric mass, $m$, which is inversely related to the square of the natural frequency, $\omega$, of the beam: $\omega \propto \sqrt{K/m}$. We hypothesize that the primary function of very high root porosity in elephant whiskers is to increase resistance to the dynamic impacts that commonly occur during fast head and trunk movements. 
Importantly, our analysis shows that this large decrease in whisker mass does not negatively impact the neuromechanics of touch in the elephant. 

More broadly, porosity is known to confer dynamic benefits to both manufactured and biological composites. Porosity allows materials to efficiently absorb energy during impact loading while simultaneously reducing mass~\cite{lazarus_review_2020}, which facilitates control and reduces the energy cost of movement. Porous roots may reduce the damage these whiskers sustain across a lifetime of mechanical use, which is essential as adult elephants spend around 16 hours per day foraging~\cite{schulz_suction_2021} and often live at least 60 years in the wild. Additionally, damage reduction is biologically crucial since elephants appear unable to regrow whiskers~\cite{deiringer_functional_2023}, unlike rodents~\cite{oliver_whisker_1966}. We conclude that porosity of the whisker root gives the elephant trunk an array of thin sensory horns that resist damage from impact while communicating clear higher-frequency signals to the mechanoreceptors in their follicles.

\subsection*{Elastomer-pointed elephant whiskers tolerate loads and encode contact location}
\begin{figure}[thbp]
      \centering      \includegraphics[width=1\textwidth,page=1,trim=0.0in 0in 0in 0.0in,clip]{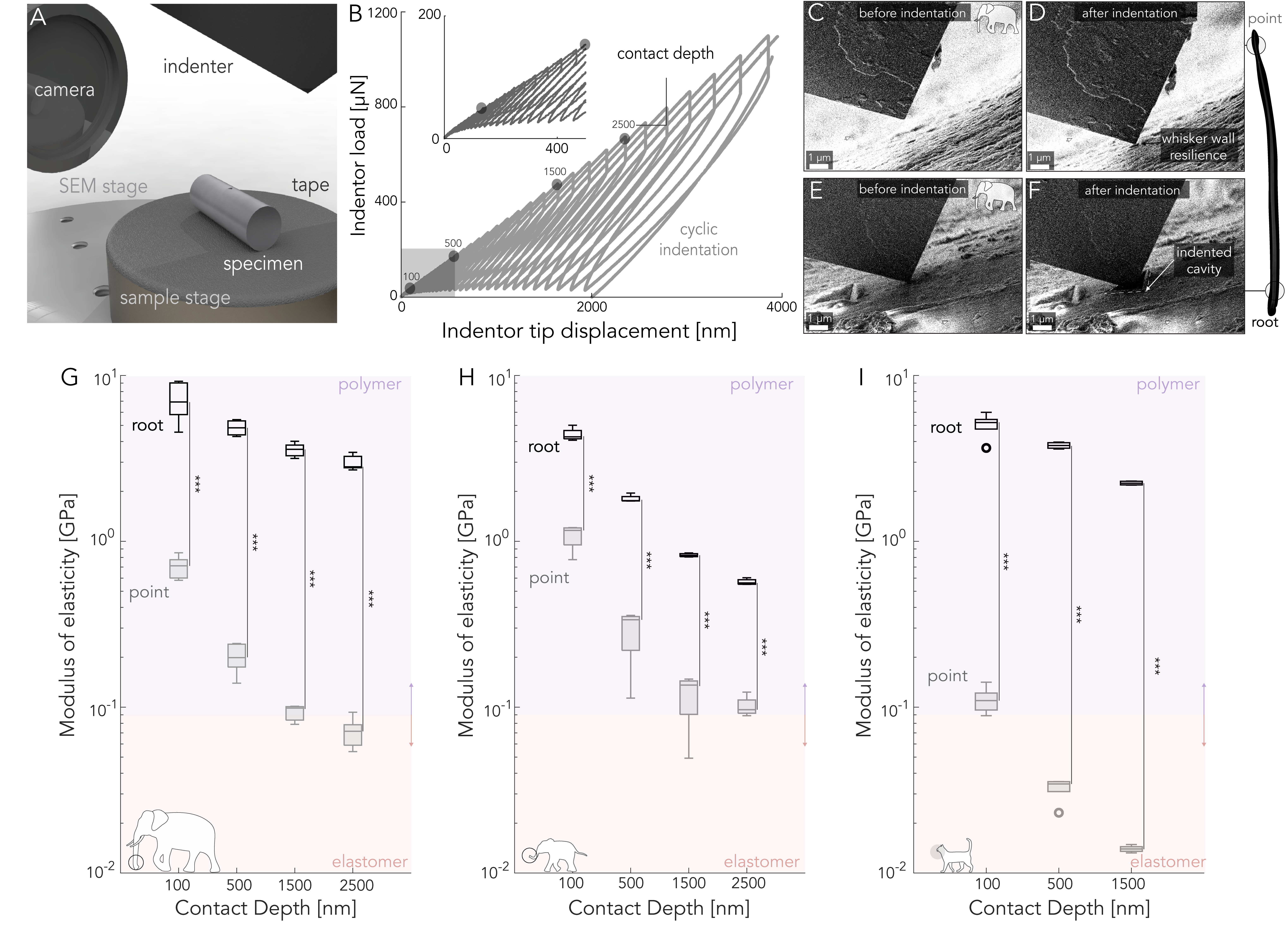}
      \caption{\label{fig3}
      \textbf{Elephant and cat whiskers have functionally graded stiffness spanning two orders of magnitude.} 
      (A) Illustration of the cube-corner indenter used for modulus characterization of whiskers. (B) Raw cyclic-loading data of cat whisker root indentation showing the specific depths at which moduli were extracted. Indentation of adult elephant whisker point (C to D) showing whisker wall resilience after indentation. Indentation of adult elephant whisker root (E to F) showing permanent plastic deformation after indentation. Indentation data shows four contact depths for the adult Asian elephant in (G), the baby Asian elephant in (H), and the domesticated cat in (I). The shaded background indicates elastomer and polymer regions~\cite{wegst_mechanical_2004}, while stars indicate significant differences ($^{\star\star\star} ~p < 0.001$). 
      }
\end{figure}

In addition to geometry and porosity, elastic modulus can also change along the length of a functionally graded beam~\cite{li_review_2020}. To determine whether the material changes along the length of elephant whiskers, we used pico-indentation into the wall at the root and at the point of whiskers from the distal section of the studied elephant trunks (\fig{fig3}A--F, fig.\ S6A-B). When an Asian elephant is two weeks old, its distal whiskers already have a significant functional material grade of almost one order of magnitude: the root modulus of $0.57\pm 0.02$ GPa ($n=4$) shifts to a softer point modulus of $0.1013\pm 0.004$\,GPa ($n=4$) $p  < 10^{-5}$, \fig{fig3}E). As the elephant develops, this functional gradient greatly increases, and the root modulus increases by an order of magnitude to $2.99 \pm 0.28$\,GPa ($n=5$). The point of the whisker changes by a comparatively small amount to $0.706\pm 0.008$ GPa ($n=5$), meaning adult elephant's whiskers have a significant stiffness gradient spanning two orders of magnitude ($p < 10^{-5}$, \fig{fig3}F), matched to that of squid's composite beak~\cite{miserez_transition_2008}. We see a similarly significant large modulus shift ($p < 10^{-5}$) in the studied cat whiskers (\fig{fig3}I) with a root modulus of $2.24 \pm 0.05$\,GPa ($n=5$) and a point modulus of $0.01 \pm 0.0005$ ($n=5$); this shifting from a stiff root to a soft point is opposite the gradient found in rats, though their modulus shifts by only a factor of 1.5 from root to point~\cite{quist_variation_2011}. All previously studied whiskers (table S3) show only a small modulus range around a factor of 3 from root to point. 

Our indentation data demonstrates that the elephant's whisker exhibits nonuniform elastic and plastic mechanical behavior at different locations along its length. Following indentation near the root, the whisker wall remains deformed (\fig{fig3}E-F), a plasticity behavior commonly observed in polymers (0.09\,GPa $\leq$ natural polymers $\leq$ 100\,GPa). Following indentation at the point, the elephant whisker responds elastically (\fig{fig3}C-D), with no permanent deformation, a behavior commonly observed in elastomers ($\leq$ 0.09\,GPa)~\cite{wegst_mechanical_2004}. Therefore, baby and adult elephant whiskers are natural polymer-elastomer composites (\fig{fig3}G-H), which have never before been observed in a mammalian keratin structure; the closest finding showed that hydration can reduce the moduli of alpha-keratin structures by one order of magnitude~\cite{greenberg_regulation_2013}.

\begin{figure}[thbp]
      \centering    \includegraphics[width=1\textwidth,page=1,trim=0.0in 0in 0in 0.0in,clip]{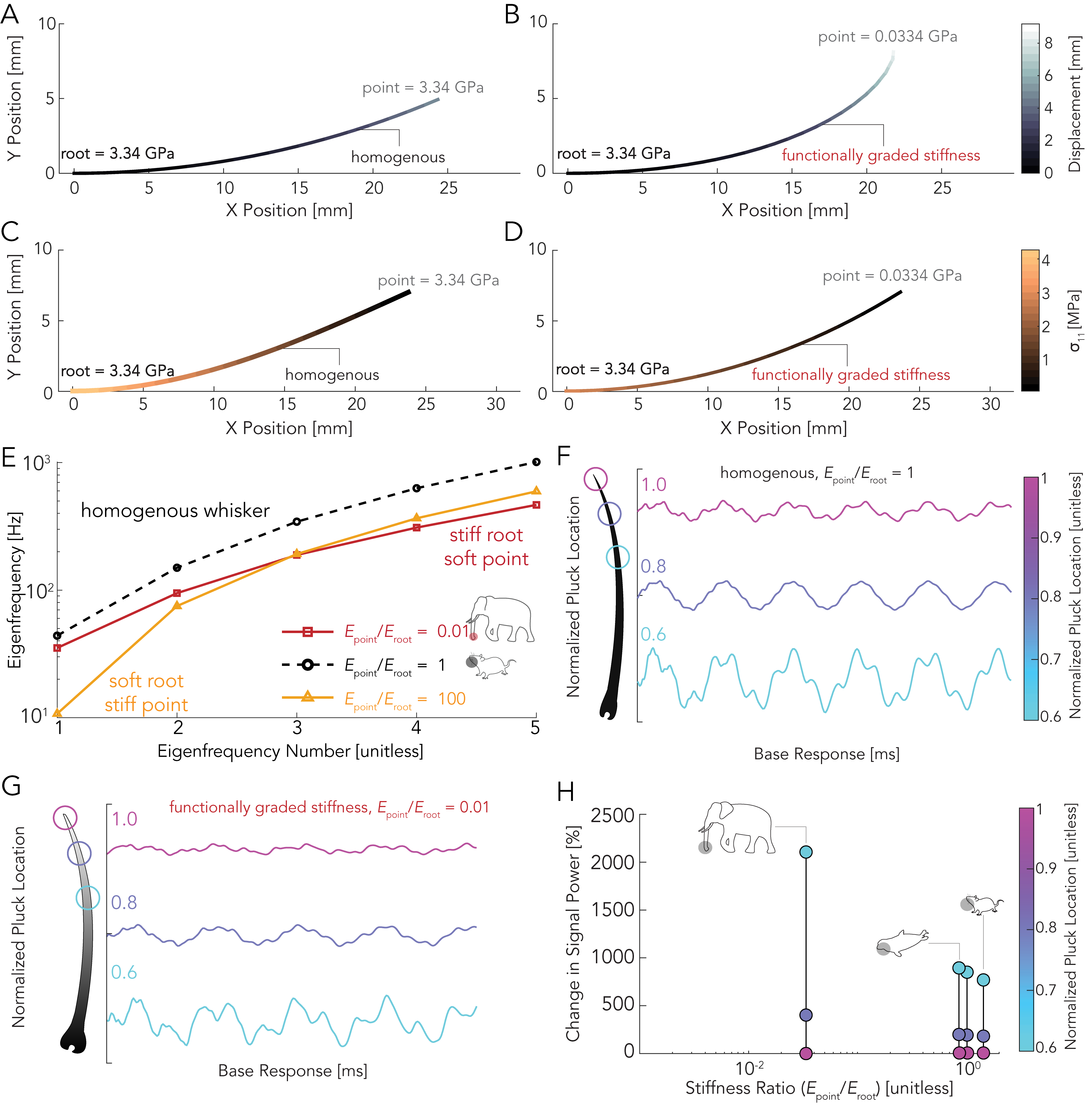}
      \caption{\label{fig4}
      \textbf{Functionally graded elastic modulus provides both static and dynamic benefits for tactile sensing.} 
      Simulation results when fixed-root whiskers are loaded with an equivalent transverse force at the point, with homogeneous in (A) and functionally graded stiffness in (B).
      Simulation results show that when whisker point displacement is prescribed, the root of a stiff ($E$ = 3\,GPa) homogeneous whisker in (C) experiences far higher stress than that of a whisker with functionally graded stiffness in (D). 
      (E) Simulated frequencies of free vibration for homogeneous stiff (black) and stiffness-gradient (red and yellow) whisker structures (fig.\ S8).
      Simulation outputs of the whisker response after transient contact for homogeneous in (F) and soft-point-to-stiff-root functionally graded stiffness in (G). 
      (H) Simulation showing the percentage of signal power change communicated to the whisker root when plucked at different locations. 
      }
\end{figure}

Static finite element analysis of whisker deformation indicates that functional grading of elastic modulus from a stiff root to a soft point distributes stresses more evenly, enables larger tip displacements, and modifies the frequency response of the whisker compared to homogeneous whiskers (fig.\ S7A). We compared the mechanical response of whiskers with the modulus gradient measured in adult elephant whiskers against uniform whiskers of constant modulus 3.3\,GPa (fig.\ S8)~\cite{quist_variation_2011} and predicted the displacement and stress fields when the root is fixed in space, and displacements or loads are prescribed at the point. Graded whiskers exhibit nearly double the displacement magnitude at equivalent point loading compared to homogeneous whiskers (\fig{fig4}A-B). Mechanoreceptors near the distal portion of the dermis in the skin would detect this increased displacement from the whisker's follicle, communicating a large deformation to the strain mechanoreceptors located at the root surrounding the whisker follicle. 


A soft point confers added benefits for interacting with rigid stationary objects such as stone, as the whisker point can lightly brush past rigid external stimuli, an advantage of soft materials~\cite{gul_fully_2018}. In simulations that reflect this scenario, where large transverse displacements at the point are prescribed (fig.\ S7B), the functionally graded elephant whisker exhibits 33\% lower peak stresses at the root relative to homogeneous whiskers (\fig{fig4}C-D, movie S2). Reduction in stress concentrations at the whisker root leads to a reduced chance of failure or breakage compared to an isotropic structure with constant properties. This reduction of stress has previously been shown in the slender structures on gecko feet, known as setae, which are functionally graded in modulus~\cite{dong_functionally_2020}. 

Whiskers with a modulus ratio $ E_{\mathrm{point}}/ E_{\mathrm{root}} \approx 10^{-2} $ from point to root, including those of the adult elephant and domesticated cat, have a first natural frequency that is double that of whiskers with the inverse ratio (\fig{fig4}E, movie S3). Whiskers with a soft root and stiff point also likely do not exist in nature as the vibration communicated to the root is so reduced in amplitude and frequency that it falls in the smallest range of sensing capable for mammalian mechanoreceptors (\fig{fig4}E). Whiskers like the elephant's (stiff root, soft point) have a slightly smaller vibrotactile signal compared to homogeneous whiskers, with their first natural frequency reduced by 15\%; however, this reduction is offset by the amplification from the porous root (\fig{fig2}J). 

Finally, we look at how modulus grading affects tactile perception of impacts along the length of the whisker. We model whiskers using nonlinear beam finite elements in a loading scenario that approximates brushing past obstacles and subsequent free vibration in space (fig.\ S7D). We monitor the time-varying reaction moment at the root of the simulated whisker, where position and orientation are fixed, and compute the discrete Fourier transform of this signal to observe the frequency content of the information transmitted to the sensory ring (fig.\ S9A). Functionally graded whiskers transmit vibrations of exponentially higher amplitude under equivalent loading compared to homogeneous stiff whiskers (\fig{fig4}F-G, fig.\ S9C, movie S4). 

As the excitation location moves toward the root of a graded whisker, the vibration amplitude transmitted to the sensory ring grows non-linearly, enabling logarithmic differentiation (\fig{fig4}G) of transient contacts along the whisker. In contrast, a whisker with uniform properties exhibits near-linear signal differentiation (\fig{fig4}F). Signal power at the whisker root, known to drive the firing of sensory neurons~\cite{lottem_mechanisms_2009}, changes much more dramatically with contact location in graded whiskers than in uniform whiskers; compared to homogeneous whiskers in identical conditions, our simulations of graded whiskers show a $2\,000\%$ increase in signal power when plucked at $60\%$ of their length (\fig{fig4}H), which is an additional 4\,dB of signal power transmitted to the root (fig.\ S9D). These functional benefits of stiffness-gradient whiskers should allow more precise localization of transient contact along the whisker length.

\subsection*{Conclusions}
Biological structures have functionally graded material properties, but these anisotropies are frequently ignored in favor of simple isotropic models that vary only in geometry. This assumption has limited our understanding of whisker-mediated touch and prevented bio-inspired whiskers from taking advantage of material shifts in porosity and modulus; existing designs are primarily soft materials that lack rigidity~\cite{sofla_haptic_2024} or rigid materials that lack flexibility~\cite{sayegh_review_2022}. This work showed that the alpha-keratin in elephant whiskers is functionally graded in geometry, porosity, and stiffness. The porosity and stiffness gradients of elephant trunk whiskers directly influence the frequency, amplitude, and power of the mechanical signals felt by mechanoreceptors in the follicle upon mechanical stimulation. Shifting from a stiff root to a soft point also amplifies the signal power connected to the firing of sensory neurons~\cite{lottem_mechanisms_2009}, potentially improving the animal's ability to perceive the location of contact along each whisker, which would aid navigation and manipulation. 

Biological functionally graded material composites like elephant whiskers can inspire engineered devices that use functional gradients to achieve specific capabilities ranging from fatigue reduction to signal power increases. One of the first animal stiffness gradients discovered was the beak of the Humboldt squid~\cite{miserez_transition_2008}, but mimicry of this stiffness gradient in soft materials posed a considerable manufacturing challenge at the time of this discovery. Recent advances in multi-material 3D printing enable unprecedented control over the deposition of materials with widely varying mechanical properties; cutting-edge inkjet systems create monolithic parts from materials with elastic moduli that span three orders of magnitude~\cite{slesarenko_towards_2018,mueller_mechanical_2015}. Recent characterization of composites built from these materials enables inverse design, whereby one achieves desired system properties such as stiffness, toughness, and frequency response by prescribing both geometry and constituent materials at the microscale~\cite{smith_digital_2024}. Fields ranging from material science and neuroscience to haptics and bio-inspired robotics rely on signal processing through material interfaces, and functional gradients have significant potential to enable programmable signal shifts tuned to specific use cases.

\subsubsection*{Acknowledgments}
The authors thank C. Schwarz for advice on whisker neuromechanics, B. Sharratt for conversations regarding functionally graded materials, T. Hildebrandt for assistance in acquisition of the elephant specimens, S. Merker and the Stuttgart State Museum of Natural History for the bighorn sheep horn sample, the Berlin Zoological Garden for providing Asian elephant adult head hair, R. Faulkner for assistance with creating physical whisker mimics, S. Griego for assistance with SEM images, B. Javot for assistance with bighorn sheep horn sample preparation, N. Reveyaz for whisker collection assistance, and J. Burns for help with data curation. L.K. thanks the Berlin School of Mind and Brain at Humboldt University. M.B. thanks the NeuroCure Cluster of Excellence at Humboldt University.

\subsubsection*{Funding:} This work was supported by the Alexander von Humboldt Foundation (to A.K.S.), the Max Planck Society (to G.R. and K.J.K.), the Carl-Schneider-Stiftung (to G.R. and K.J.K.), BCCN Berlin, Humboldt-Universität zu Berlin (to L.V.K. and M.B.), and the Deutsche Forschungsgemeinschaft (DFG, German Research Foundation) under Germany's Excellence Strategy – EXC-2049 – 390688087 (to L.V.K. and M.B). 

\subsubsection*{Author Contributions:} A.K.S., G.R., and K.J.K. conceived the idea and designed the experiments. A.K.S., M.B., G.R., and K.J.K. supervised the research project. A.K.S., L.V.K., L.T.S., D.S.P., H.D., J.L., and G.R. conducted the experiments. A.K.S., L.V.K., L.T.S., D.S.P., H.D., J.L., M.B., G.R., and K.J.K. analyzed the data. All authors contributed to the writing and editing of the manuscript. 

\subsubsection*{Competing interests:}
The authors declare no competing interests.

\subsubsection*{Data and materials availability:} Raw indentation data, microCT data, and SEM images can be found at~\cite{schulz_data_2025}. The authors have provided bitmap prints of all SEM images in the data repository for digital accessibility, along with lithograph 3D model files made using the process from Koone et al.~\cite{Koone22-SA-Graphics}. The authors have also provided a repository for finite element simulation of whiskers with stiffness, porosity, and/or geometry gradients in static and dynamic cases~\cite{smith_whiskeranalyses_2025}. 

\newpage
\singlespacing
\scriptsize
\begin{footnotesize}

\end{footnotesize}
\bibliographystyle{unsrt}

\end{document}